\documentclass[a4paper,fleqn,usenatbib]{mnras}

\usepackage{mathptmx}
\usepackage[T1]{fontenc}
\usepackage{ae,aecompl}

\usepackage{graphicx}	% Including figure files
\usepackage{amsmath}	% Advanced maths commands
\usepackage{amssymb}	% Extra maths symbols

\title[phase transition in magnetized neutron stars]{Effects of the quark-hadron phase transition on highly magnetized neutron stars}

\author[B. Franzon et al.]{
B. Franzon,$^{1}$\thanks{franzon@fias.uni-frankfurt.de}
R.O. Gomes,$^{2}$\thanks{rosana.gomes@ufrgs.br}
S. Schramm$^{3}$ \thanks{schramm@fias.uni-frankfurt.de}
\\
$^{1}$Frankfurt Institute for Advanced Studies,
Ruth-Moufang - 1 60438, Frankfurt am Main,
Germany\\
$^{2}$Instituto de F\'isica, Universidade Federal do Rio Grande do Sul, Porto Alegre-RS 91501-970, Brazil\\
$^{3}$Frankfurt Institute for Advanced Studies,
Ruth-Moufang - 1 60438, Frankfurt am Main,
Germany
}

\date{Accepted XXX. Received YYY; in original form ZZZ}

\pubyear{2016}

\begin{document}
\label{firstpage}
\pagerange{\pageref{firstpage}--\pageref{lastpage}}
\maketitle

\begin{abstract}
The presence of quark-hadron  phase transitions in neutron stars can be related to several interesting phenomena. In particular, previous calculations have shown that fast rotating neutron stars, when subjected to a quark-hadron phase transition in their interiors, could give rise to the backbending phenomenon characterized by a spin-up era. In this work, we use an equation of state composed of two phases, containing nucleons (and leptons) and quarks. The hadronic phase is described in a relativistic mean field formalism that takes many-body forces into account, and the quark phase is described by the MIT bag model with a vector interaction. Stationary and axi-symmetric stellar models are obtained in a self-consistent way by solving numerically the Einstein-Maxwell equations by means of a pseudo-spectral method.   As a result, we obtain the interesting backbending phenomenon for fast spinning neutron stars. More importantly, we show that a magnetic field, which is assumed to be axi-symmetric and poloidal, can also be enhanced due to the phase transition from normal hadronic matter to quark matter on highly magnetized neutron stars. Therefore, in parallel to the spin-up era, classes of neutron stars endowed with strong magnetic fields may go through a ``magnetic-up era" in their lives. 
 
\end{abstract}

\begin{keywords}
equation of state -- magnetic fields -- neutron stars
\end{keywords}

%%%%%%%%%%%%%%%%%%%%%%%%%%%%%%%%%%%%%%%%%%%%%%%%%%

%%%%%%%%%%%%%%%%% BODY OF PAPER %%%%%%%%%%%%%%%%%%

\section{Introduction}
Compact stars refer collectively to white dwarfs (WD), neutron stars (NS), or even black holes (BH). The latter two are usually formed in catastrophic astrophysical events such as supernova explosions, see e.g.  \cite{glendenning2012compact, shapiro2008black}. In a few seconds, these explosions release the brightness of millions of suns, so it is not surprising that such phenomena have been observed since ancient times. The kind of the remnant depends primarily on the mass of the proto-star. Moreover, these objects harbour compressed ultra-dense matter in their interiors, have spin rotation from zero to milliseconds and can support ultra-high magnetic fields. These features, in
conjunction  with the corresponding progress in observational astrophysics, make them one of the most and, sometimes, the only suitable environment to inspect, or at least to infer from, the behaviour of matter and electromagnetic fields under extreme conditions.

Depending on the equation of state (EoS) chosen to describe the matter, the density inside NS can reach values several times higher than the nuclear saturation density $\rho_{0}\sim0.15\,\rm{fm^{-3}}$. In fact, during the past years, much effort has been made in order to describe and shed light on the still open question concerning the equation of state for ultra-dense nuclear matter.  Furthermore, another important question that we address in this work is the impact of the magnetic field on the structure and on the evolution of these stars.  The knowledge of the microscopic theory related to the equation of state allows not only to study the general structure of the star, including its deformation due to the magnetic field and due to rotation, but also provides information about the inner composition of the star, including possible exotic phases such as quarks and hyperon in the stellar core.

As shown in \cite{glendenning1997signal, zdunik2006phase}, phase transitions inside neutron stars can be accessed through the backbending phenomenon, where stars spin up over time as a consequence of a phase change in their cores. However, the exact mechanism responsible for the spin-down of stars is still unclear. The most accepted idea is that these stars spin down because of magnetic torques and lose energy through magnetic dipole radiation (MDR).  In this way, rotating magnetized stars behave as  oblique rotators \cite{pacini1967energy, pacini1968rotating, gold1975rotating}. In addition to the dipole radiation, processes such as emission of gravitational radiation and pulsar wind can also contribute to the braking index of neutron stars \cite{ostriker1969nature,ferrari1969theoretical,blandford1988interpretation, manchester1985second}. In all cases, the star energy loss can be described by a power law $\dot{E}=-C\Omega^{n+1} $, being $C$ a term that accounts for the pulsar structure, $E$ for the kinetic energy, $\Omega$ is the pulsar angular velocity and $n$ is the braking index, which, therefore, describes the dependence of the braking torque on the rotation frequency. 
 
According to the MDR theory, a rigid star with a constant dipole magnetic field and a constant moment of inertia $I$ has the canonical braking index of $n = 3$. As calculated in \cite{pacini1967energy, pacini1968rotating}, the spin-down relation for such stars is given by $\dot{\Omega} \propto \Omega{^3}$. However, in the presence of rotation, the moment of inertia $I$ is not constant in time and, therefore, one has to considerer the dependence $I(\Omega)$. In this case, the star responds to changes due to the centrifugal force becoming oblate. This effect reduces the value of the braking index from the standard oblique rotator model to a value of $n<3$, see \cite{Hamil:2015hqa}.  Furthermore, the microphysics impacts strongly the braking index, which can have values in the range of $-\infty<n<\infty$ in presence of a first-order phase transition, see e.g. \cite{glendenning1997signal, chubarian1999deconfinement}.

In case of a phase transition, its duration epoch is governed by a slower loss of the star angular momentum due to radiation. This will be seen in the behaviour of the $I(\Omega)$ curve presented in section III.  Similar results as those presented in section III were already investigated before, see for example \cite{glendenning1997signal, zdunik2006phase, chubarian1999deconfinement}.  At the same time, as we will see in section IV, this phase change in the stellar interior can introduce not only a spin-up era, but also an increase in the magnetic field throughout the star.

From pulsar spin-down observations together with the magnetic dipole model, the magnitude of the surface magnetic field in NS is typically estimated to be of the order of $10^{12}$-$10^{13}$ G.  However,  other classes of neutron stars known as Anomalous X-ray Pulsars and Soft-Gamma-Ray-Repeater, referred to magnetars,  can have surface magnetic fields as large as $10^{14}$-$10^{15}$ G \cite{Duncan:1992hi,Thompson:1993hn, Thompson:1996pe,  paczynski1992gb, melatos1999bumpy}. Such strong magnetic fields affect both the structure and the composition of these stars and can, potentially, convert a hybrid star into a hadronic star as shown in \cite{Franzon:2015sya}.   

By modelling rapidly rotating neutron stars, we show the effects of a phase transition on the moment of inertia $I(\Omega)$ of the stars and, as a consequence, the impact on the braking index $n(\Omega)$  (which becomes frequency-dependent). In addition, we constructed models for highly magnetized neutron stars and we show that, similarly to the rotating case, the magnetic field induces a reduction in the moment of inertia, which introduces an era in the stellar evolution where the magnetic field increases.  In comparison with the spin-up era for fast rotating bodies, we call this phenomenon a ``magnetic-up era".

\section{Equation of state}
The transition from confined to deconfined matter inside neutron stars has been extensively studied over the last years and, in many cases, also applied to study properties of hybrid stars 
\cite{Bombaci:2006cs, Bombaci:2009jt, Yasutake:2010eq, Dexheimer:2012mm, Lenzi:2012xz, Ayvazyan:2013cva, Brillante:2014lwa, 
Alvarez-Castillo:2014dva, Dexheimer:2014pea, Franzon:2015sya, deCarvalho:2015lpa}. 
Despite this progress, at the present moment, there are still substantial uncertainties in the equation of state  regarding the description of the stellar matter at  supra-nuclear densities. 

In this work, we use an equation of state with quark-hadron phase transition to describe the stellar interior. The hadronic phase, composed of nucleons (together with leptons), is described in the framework of a relativistic  mean-field theory  and takes into account many-body forces contributions in the baryon couplings. For the quark phase, we use the MIT bag model with vector interaction, in order to reproduce the maximum effect that phase transitions can produce. We choose in this work to describe the deconfinement phase transition by using the usual Maxwell construction, i.e, reproduce a sharp phase transition. 
%Furthermore, depending on the surface tension among the phases, the first phase transion can occur
%in either a sharp transition, with constant pressure and a discontinuous density (Maxwell construction),
%or in a continuous scenario, with the appearance of a mixed phase (Gibbs construction).

%In this work, we describe hybrid stars by employing an equation of state  with two distinguished phases that are connected by a Maxwell construction in a sharp phase transition. The hadronic phase is described by the many-body forces (MBF) formalism \cite{Gomes:2014aka}, in which the interaction among baryons is mediated by the exchange of mesons, in a mean-field approximation. The main feature of this model is the introduction of a field dependence of the scalar coupling constants
%that simulates the influence of many-body forces as a medium effect. 
%This formalism has been successfully used to determine the nuclear matter properties at saturation, as well as massive hyperon stars.

%For the purpose of emphasizing the effects of the phase transition in our calculations, we neglect the hyperonic degrees of freedom, since hyperons may also play an important role in hybrid stars, see e.g. \cite{zdunik2004hyperon}.
%For describing the quark phase, we adopt the MIT bag model with vector interaction.
%This interaction is introduced in order to add repulsion to the EoS at high densities and, therefore, to be able to describe massive hybrid stars \cite{Gomes:2015fsa}.

\subsection{HADRONIC PHASE}\label{Hphase} 
The lagrangian density of the MBF model reads:

\small
\begin{equation}\begin{split}\label{lagrangian}
\mathcal{L}&= \underset{b}{\sum}\overline{\psi}_{b}\left[\gamma_{\mu}\left(i\partial^{\mu}-g_{\omega b}\omega^{\mu} 
-g_{\varrho b}\mathbf{\textrm{\ensuremath{I_{3b}}\ensuremath{\varrho_3^{\mu}}}}\right)
-m^*_{b \zeta}\right]\psi_{b}
\\& +\left(\frac{1}{2}\partial_{\mu}\sigma\partial^{\mu}\sigma-m_{\sigma}^{2}\sigma^{2}\right)
+\frac{1}{2}\left(-\frac{1}{2}\omega_{\mu\nu}\omega^{\mu\nu}+m_{\omega}^{2}\omega_{\mu}\omega^{\mu}\right)
\\&+\frac{1}{2}\left(-\frac{1}{2}\boldsymbol{\varrho_{\mu\nu}.\varrho^{\mu\nu}}+m_{\varrho}^{2}\boldsymbol{\varrho_{\mu}.\varrho^{\mu}}\right)
+\left(\frac{1}{2}\partial_{\mu}\boldsymbol{\delta.}\partial^{\mu}\boldsymbol{\delta}-m_{\delta}^{2}\boldsymbol{\delta}^{2}\right)
\\& +\underset{l}{\sum}\overline{\psi}_{l}\gamma_{\mu}\left(i\partial^{\mu}-m_{l}\right)\psi_{l}.
\end{split}\end{equation}
\normalsize

The subscripts $b$ and $l$ correspond to nucleonic ($n$, $p$) and lepton ($e^-$, $\mu^-$) degrees of freedom. The first and last terms in Eq.~\ref{lagrangian} represent the Dirac lagrangian density for nucleons and leptons. The other terms represent the lagrangian densities of the scalar mesons $\sigma$ and $\delta$ (Klein-Gordon lagrangian density), and the vector mesons $\omega$ and $\varrho$ (Proca lagrangian density).  The meson-baryon interaction appears in the first term, contained in the coupling constants ($g_{\omega b}$,$g_{\varrho b}$) and effective masses ($m^ {*}_{b \zeta}$). The properties of the particles used in this work can be found in the tables (\ref{particles}) and (\ref{campos}).
In this formalism, the $\delta$ and $\varrho$ fields allow for a better description of the isospin asymmetry of the system. See \cite{Alaverdyan:2010zz} for an analyses of the influence of the delta meson on phase transitions.

The effects of the many-body forces contribution (controlled by the $\zeta$ parameter) and the nuclear interaction 
on the baryonic effective masses and chemical potentials are expressed in the following:

\begin{equation}
m_b^* = m_b -\left(1+    \frac{g_{\sigma b}\sigma+ g_{\delta b}I_{3b}\delta_{3}}{\zeta m_{b}}  \right)^{-\zeta} \left(  g_{\sigma b}\sigma + g_{\delta b}I_{3b}\delta_{3} \right)
\label{meff}
\end{equation}

\begin{equation}\label{mu_eff}
 \mu^*_{b_i}= \sqrt{k_{f_{b}}^2+(m_{b \zeta}^{*})^2} + g_{\omega b}\omega +g_{\varrho b}  I_{3b} \varrho_{3} + g_{\phi b}\phi,
\end{equation}
where $k_{f_b}$ and $m_b$ correspond to the fermi momenta and the bare masses of the baryons.

By choosing small values of the $\zeta$ parameter, the second term in Eq.~\ref{meff} can be expanded by means of 
nonlinear self-couplings of the scalar fields ($\sigma$, $\delta$), simulating the effects of 
many-body forces in the nuclear interaction.  
Each set of parameters generates different EoS's and, hence, different sets of nuclear saturation properties.
For this study, we use the parameterization: $\zeta=0.040$, reproducing a binding energy of $B/A = -15.75\,$MeV and 
the saturation density $\rho_0=0.149\,\,\rm{fm^{-3}}$. This set of parameters reproduces the following nuclear saturation properties: 
effective mass of the nucleon $m^{*}/m=0.66$, incompressibility $K_0=297\,$MeV, 
symmetry energy $J_0=32\,$MeV and slope of the symmetry energy $L_0=97\,$MeV
(for more details of nuclear saturation properties covered by the MBF model, see e.g. \cite{Gomes:2014aka}).

\begin{table}
\caption{Properties of the particles used in the formalism: mass $m$, projection of the isospin in the $z$ direction $I^{3}$, baryon charge $q_{b}$ and electric charge $q_{e}$.} 
\begin{center} \label{particles}
\begin{tabular}{|c|c|c|c|c|c|c|c|} 
\hline 
Particle & Mass $(MeV)$ & $I^{3}$ & $q_b$ & $q_e$ \tabularnewline
\hline
\hline
$p$ & $939.6$ & $1/2$ & $1$ & $+1$ \tabularnewline
\hline
$n$ & $938.3$ & $-1/2$ & $1$ & $0$  \tabularnewline
\hline
$e^{-}$ & $0.511$ & $0$ & $0$ & $-1$  \tabularnewline
\hline
$\mu^{-}$ & $105.7$ & $0$ & $0$  & $-1$ \tabularnewline
\hline
$u$ & $1.0$ & $-$ & $1/3$ & $+2/3$  \tabularnewline
\hline
$d$ & $1.0$ & $-$ & $1/3$ & $-1/3$  \tabularnewline
\hline
$s$ & $100.0$ & $-$ & $1/3$  & $-1/3$ \tabularnewline
\hline\hline
\end{tabular}
\par\end{center}
\end{table}

\begin{table}
\caption{Meson fields properties considered in the hadronic phase and their respective coupling 
constants at nuclear saturation density.} 
\begin{center} \label{campos}
\begin{tabular}{|c|c|c|c|c|} 
\hline 
Meson & Classification & Mass & $(g_{i_N}/m_i)^ 2$ \\
& &  (MeV) &  $ (fm^ 2) $  \\ 
\hline \hline
$\sigma $ & scalar-isoscalar & 550 & 14.51 \tabularnewline
$\mbox{\boldmath$\delta$}$  & scalar-isovector & 980 & 0.38 \tabularnewline
$\omega_\mu $ &  vector-isoscalar & 782 & 8.74 \tabularnewline
$\varrho $  & vector-isovector & 770 & 4.47 \tabularnewline
\hline\hline
\end{tabular}
\end{center}
\end{table}

\subsection{QUARK PHASE}\label{Qphase} 
For the quark phase, we adopt the MIT bag model with an additional vector interaction. We take the quarks $u$, $d$ and $s$, 
and the leptons $e^-$ and $\mu^-$  as degrees of freedom (see Table \ref{particles}). 
The lagrangian density of the model reads:

\small
\begin{align}
\mathcal{L} = & \underset{q}{\sum}\left[ \overline{\psi}_{q}  (i\gamma_{\mu}\partial^{\mu}-g_{V q}\gamma_{\mu}V^{\mu} -m_q - B)\psi_{q}\right]\theta_H  \nonumber\\
+ & \underset{l}{\sum}\overline{\psi}_{l}\gamma_{\mu}\left(i\partial^{\mu}-m_{l}\right)\psi_{l} ,
\label{lagrangianQ}
\end{align}
\normalsize
with $\theta_H$ being the Heavyside function responsible for the confinement feature of the model ($\theta_H = 1$ inside the bag; $\theta_H = 0$ outside), 
and $B$ is the bag constant. This constant represents the extra energy per unit of volume required to create a region of perturbative vacuum \cite{Farhi:1984qu}. 

The vector interaction is introduced by a vector-isoscalar meson $V^\mu$,  with coupling constant $g_V$, coupling to all three quarks.
As discussed in \cite{Shao:2013toa}, the field that introduces the vector interaction can be considered analogous to the $\omega$ field 
used in the hadronic model.
As in the hadronic phase, the vector interaction introduces a shift in the quark chemical potential:
\begin{equation}\label{mu_eff}
 \mu^*_{q_i}= \sqrt{k_{f_{q}}^2+m_q^2} + g_{V q}V.
\end{equation}

The vector interaction has been extensively studied in the context of bag models \cite{contrera2014hadron,klahn2015vector} and  Nambu-Jona-Lasinio models \cite{contrera2014hadron,klahn2015vector, ranea2015constant}, 
and has been applied to different investigations such as  the study of the phase structure and transitions of matter \cite{Shao:2012tu, contrera2014hadron}, the effect of strong magnetic fields \cite{denke2013influence,menezes2014repulsive}, 
the thermal evolution of neutron stars \cite{deCarvalho:2015lpa} and others.
In particular, the approach used in this work is analogous to the one proposed in \cite{Alford:2004pf,Weissenborn:2011qu}, 
which introduces phenomenological corrections based on gluon effects that result in extra terms in the EoS and particles densities of quark matter. 
 Note that the introduction of a vector-isovector field generates these same shifts in the EoS, due to the additional term, $V^{\mu}$, in the pressure, as well as in the energy density. In addition, $V^{\mu}$ changes the particle population, as the quark chemical potentials are modified.

In this context, the hardest task of such models is to determine the coupling constant of the vector interaction.
There are attempts to account for the interactions of quarks and gluons and, then, 
constrain the value of the coupling by incorporating higher orders of perturbation theory 
and radiative corrections as done in \cite{Fraga:2001id,Fraga:2013qra,Restrepo:2014fna}. 
However, such couplings remain widely uncertain and may also 
have a dependence on the density and temperature.
 
%i.e., a dependence on density and temperature (SHOULD WE KEEP THIS PHRASE?). 
% mention comparison to lattice QCD results?
%**STABILITY OF MATTER (ADDRESS THE CORRESPONDING BAG VALUE TO HAVE A STABLE QUARK MATTER) 
%(BINDING ENERGY>934MEV <??? (SURFACE - CITE ALFORD, JURGEN)
%DISCUSS THE CASE OF UNSTABLE QUARK MATTER IN ORDER TO HAVE HYBRID STARS

To conclude, we take the values of the vector coupling and bag constant to be: 
$(g_V/m_V)^2 = 2.2\,\rm{fm^2}$, $B^{1/4}=160\,$MeV, in order to reproduce massive and stable hybrid stars.
The choice of parameters of both phases permits the complete EoS to describe the symmetric nuclear matter in terms of a pure 
hadronic phase at low densities, and the regime of high density  reached in the inner core of neutron stars to be described by a quark phase.

\subsection{PHASE TRANSITION}\label{maxwell_section} 
In this section, we describe the hadron-quark phase transition in the interior of hybrid stars by using a Maxwell construction. In this case, both phases are charge neutral and the conditions on the chemical equilibrium determine the coexistence phase. This transition is described by a regime of constant pressure, which leads to a discontinuity in the 
energy density and in the baryon number density. The Maxwell criteria read:

\begin{equation}\label{maxwell}
P_H = P_Q , \qquad \mu_n^ {H}= \mu_n^ {Q}.
\end{equation}

\begin{figure}
\begin{center}
\includegraphics[width=0.7\textwidth,angle=-90,scale=0.5]{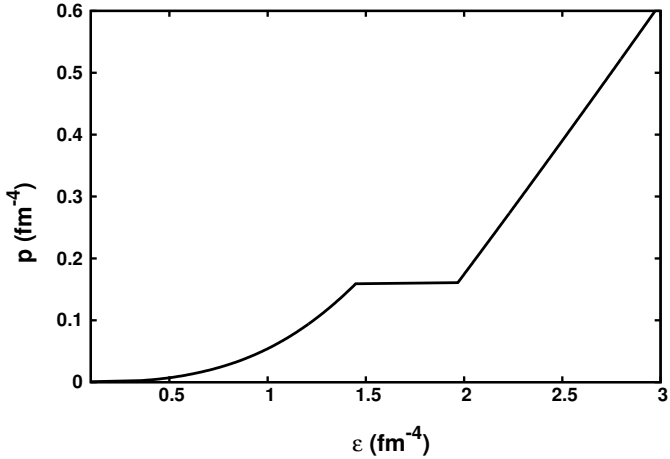}
\caption{\label{eos_plot} EOS for neutron star matter including a deconfinement phase transition for the parametrization $\zeta=0.040$ of the MBF formalism. The
bag constant is $B=$ 160 $\rm{MeV^{4}}$ and the vector coupling   $(g_V/m_V)^2$ = 2.2 $\rm{fm^2}$.}
\end{center}
\end {figure}

For the parametrizations used in this work, we found that the phase transition occurs at $\mu_n=1101.3\,$MeV, which corresponds to 
a transition pressure of $P_0=0.16\,\rm{fm^{-4}}$ and an energy gap of $\Delta \varepsilon = 0.52 \, \rm{fm^{-4}} $. The EoS including both phases is depicted in Fig.~\ref{eos_plot}.

The issue of whether the phase transition takes place in a Maxwell or a Gibbs scenario depends on the surface tension between the two phases. The comparison of both scenarios in the investigation of hybrid 
stars has been studied in several works, see e.g. \cite{Bhattacharyya:2009fg,Hempel:2009vp,Yasutake:2009kj,Yasutake:2010eq,Alaverdyan:2010zz}. 
Furthermore, the threshold values of the surface tension necessary to 
describe each type of transition scenario have been calculated in \cite{Lugones:2013ema,Garcia:2013eaa}. 
However, such estimations are highly model dependent, and the issue of possible phase transition scenarios remains an open question.

%\section{Numerical calculations}
\section{Rotating hybrid stars}
Rapidly rotating general relativistic stars were already described in \cite{komatsu1989rapidly, cook1992spin, stergioulas1994comparing}.  In this section,  we calculate stationary equilibrium configurations of  uniformly rotating cold neutron stars within a general relativity framework. We solve the Einstein field equations for stationary axi-symmetric spacetime using the C++ class library for numerical relativity LORENE (http://www.lorene.obspm.fr). The formalism, the relevant equations, the numerical procedure and tests used to construct the stellar models can be found in \cite{gourgoulhon20123+, Bonazzola:1993zz, Bocquet:1995je, Chatterjee:2014qsa}. Recently, this formalism was applied  to magnetized hybrid stars and magnetized and fast rotating white dwarfs, see e.g. \cite{Franzon:2015sya,Franzon:2015gda}.

%In the following section, effects of strong magnetic fields were also taken into consideration by solving the coupled Maxwell-Einstein equation in a self-consistent way as in Refs.~\cite{Bocquet:1995je,Chatterjee:2014qsa,Franzon:2015sya}. 

The  internal composition  of  rotating neutron stars is  modelled by the equation of state as described in section II.   The dependence of the internal structure of the NS with rotation is crucial, since the centrifugal force due to the rotation will help to stabilize the star against collapse and the star will be deformed: compressed in the polar  direction and expands in the equatorial direction.  With this in mind, different rotation frequencies produce different relations between the mass and the radius for rapidly rotating stars as shown in \cite{friedman1992rapidly, spyrou2002spin,zdunik2008strong, haensel2009keplerian,zdunik2004hyperon}.  In addition, similar calculations were done with a broad set of realistic equations of state in \cite{salgado1994high} .

\begin{figure}
\begin{center}
\includegraphics[width=0.7\textwidth,angle=-90,scale=0.5]{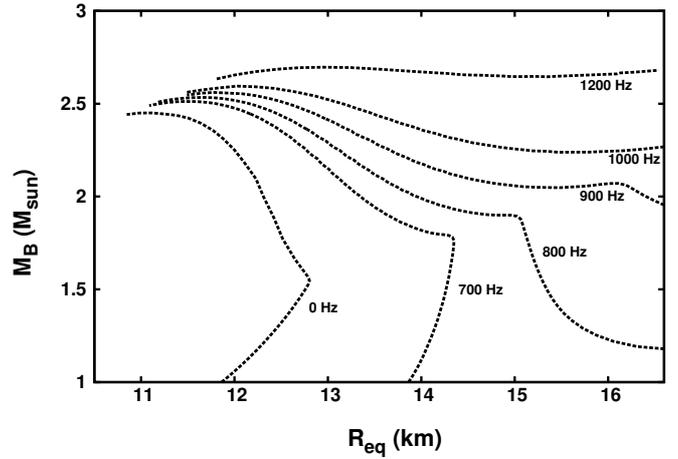}
\caption{\label{mass_radius_rotation} Baryon  mass  versus  equatorial  radius for neutron stars at different fixed rotation frequencies. These configurations were calculated using the EoS presented in Fig.~\ref{eos_plot}. }
\end{center}
\end {figure}

Effects of rotation on the backbending phenomenon in neutron stars were considered before in \cite{zdunik2006phase, chubarian1999deconfinement, cheng2002phase, heiselberg1998phase}.  As the stars spin down due to the loss of angular momentum, the central density increases and a phase transition to pure quark matter might occur \cite{cheng2002phase, glendenning1997signal}.  In order to investigate if our equation of state produces similar mass-radius diagrams as in \cite{zdunik2006phase}, in Fig.~\ref{mass_radius_rotation} we show the baryonic mass as a function of the circular equatorial radius for neutron stars with frequencies ranging from 0 to 1200 Hz. For similar mass-radius diagrams  $M_{B}(R_{eq})$, see also \cite{zdunik2004hyperon}. In this case, it was shown that equations of state with hyperon degrees of freedom can also produce the backbending phenomenon. In our case, we have neglected additional exotic phases with hyperons in order to investigate exclusively the effects of a quark-hadron phase transition.
 
 According to Fig.~\ref{mass_radius_rotation}, the mass-radius curves present inflexion points (with a minimum in the baryon mass $M_{B}$) for frequencies between $\sim$900 Hz and $\sim$1200 Hz. This defines a region (at fixed baryon masses) where the backbending phenomenon may appear in the stars. Conclusions along this line were already studied in \cite{zdunik2006phase}. The diagram in Fig.~\ref{mass_radius_rotation}  shows that, as the frequency increases, both the baryon mass and the radius of the stars increase, which is a direct effect of the centrifugal forces due to rotation. We choose to show in Fig.~\ref{mass_radius_rotation} the baryon mass instead the usual gravitational mass since this is the fixed quantity during the evolution of isolated neutron stars.
 
% We choose to show the baryon mass in Fig.~\ref{mass_radius_rotation}, because during the evolution of isolated neutron stars the baryon mass is assumed to remain constant. 

\begin{figure}
\begin{center}
\includegraphics[width=0.7\textwidth,angle=-90,scale=0.5]{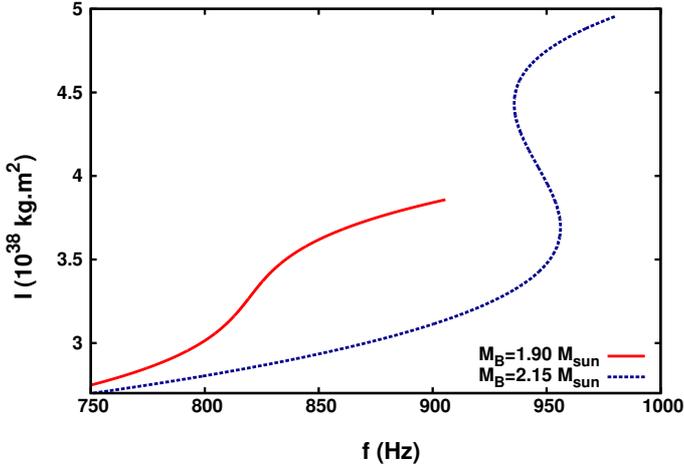}
\caption{\label{iner_omega_maxwell} Relation between the moment of inertia and the frequency for rotating neutron stars at different fixed  baryon masses, 1.90 $\rm{M_{\odot}}$ and 2.15 $\rm{M_{\odot}}$, respectively. }
\end{center}
\end {figure}

 In Fig.~\ref{iner_omega_maxwell}, we present the moment of inertia as a function of the rotational frequency $f$  for two different stars at fixed baryon masses of $M_{B}=1.90\,\rm{M_{\odot}}$ and $M_{B}=2.15\,\rm{M_{\odot}}$. Both the moment of inertia $I$ and the angular velocity $\Omega$ are decreasing functions of time. According to the Figure \ref{iner_omega_maxwell}, for the star with  $M_{B}=1.90\,\rm{M_{\odot}}$, there is a reduction in the spin-down rate when this star undergoes a phase transition. This effect is more pronounced in the case of $M_{B}=2.15\,\rm{M_{\odot}}$, in which the quark-hadron phase transition induces a spin-up era in the star's evolution. During this time,  the star will lose energy due to dipole radiation but, still, it will spin faster and grow in size.  This same effect was already reported in \cite{glendenning1997signal, weber1997signal}.      

The spin-down relation of a pulsar can be written as $\dot{\Omega}= -C{\Omega}^{n+1}$, where $C$ is a term related to the structure of the star and $n$ is the braking index, which can be obtained directly from the frequency $\Omega$ of the pulsar and its time derivatives, $\dot{\Omega}$ and  $\ddot{\Omega}$. The  braking  index can be expressed through the relation \cite{gao2015constraining}:
\begin{equation}
n=\frac{\Omega \ddot{\Omega}}{\dot{\Omega}^2}. 
\label{brak}
\end{equation}

The energy loss due to the emission of radiation can be represented by the equation:
\begin{equation}
\frac{dE}{dt}= \frac{d}{dt}\left (\frac{1}{2}I\Omega^2 \right ) = -C{\Omega}^{n+1}.
 \label{energyloss}
\end{equation}

From equation Eq.~\ref{energyloss}, one can rewrite Eq.~\ref{brak} in the form:
\begin{equation}
n(\Omega)=3-\frac{3I^{\prime}\Omega + I^{\prime\prime}\Omega^2 }{2I+I^{\prime}\Omega }, 
\label{brakfinal}
\end{equation}
with $I^{\prime}$ and $I^{\prime\prime} $ being the first and the second derivatives of the angular momentum with respect to the angular velocity $\Omega$, $\frac{dI}{d\Omega}$ and $\frac{d^{2}I}{d^{2}\Omega}$, respectively. As a result, the braking index is now written in a frequency-dependent manner, $n(\Omega)$.  

If one ignores the changes in the moment of inertia during the spin-down evolution,  it can be seen  from Eq.~\ref{brakfinal} that a purely dipole radiation yields a braking index of 3.  However, few measurements of braking index of isolated neutron stars are available in the literature (see e.g. \cite{gao2015constraining, Hamil:2015hqa} and references therein), and in all cases, one has $n<3$.  In order to determine accurately the braking index $n$, it is necessary high precision measurements of the angular velocity $\Omega$ and its corresponding time derivatives $\dot{\Omega}$, which show how stars are slowing down. For this reason, braking index observations are much easier for young pulsars, not only because they spin very fast, but also because the braking is not affected by low timing noise or glitches.  In addition, for older pulsars, the measurements of $\dot{\Omega}$ and $\ddot{\Omega}$ might require many years and yield very small values.         

\begin{figure}
\begin{center}\includegraphics[width=0.7\textwidth,angle=-90,scale=0.5]{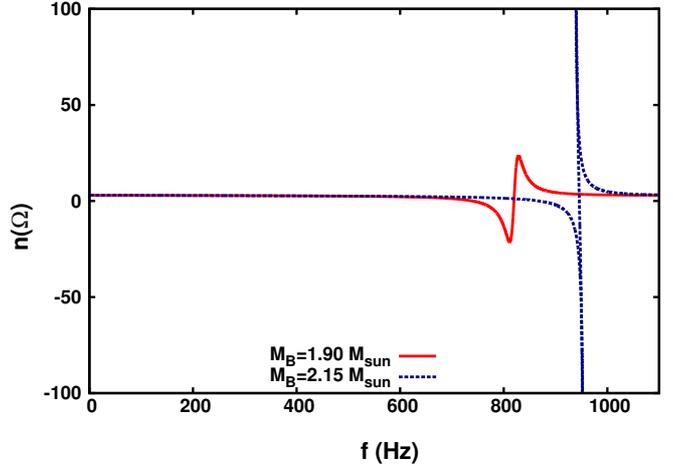}
\caption{\label{brak_maxwell} Braking index as a function of the frequency for rotating neutron stars. The curves correspond to the same stars as shown in Fig. \ref{iner_omega_maxwell}}.  
\end{center}
\end {figure}

In order to evaluate the braking index in presence of a quark-hadron phase transition, we make use of the rotating configurations already shown in Fig.~\ref{iner_omega_maxwell}. The results are depicted in Fig.~\ref{brak_maxwell}.  
In this case, the braking index does not deviate from 3 for slow rotation. However, as the frequency increases, it can reach values far from 3. In addition, when the phase transition is reached in the core of the star, there is an anomalous behaviour in the braking index curve $n(\Omega)$, whose extreme case is seen for the higher mass case $M_{B}=2.15\,\rm{M_{\odot}}$, where the braking index reaches values from $-\infty$ to $+\infty$.

Besides rotation, strong surface magnetic fields can be observed in some NSs, with values up to $10^{15}$ G. In this case, one expects to find even stronger magnetic fields inside these stars. For example, according to the virial theorem, compact stars can support internal magnetic fields of up to $10^{18-20}$ G (see e.g. \cite{ferrer2010equation, lai1991cold, fushiki1989surface, cardall2001effects}), turning such stars into a distinguished species than conventional radio pulsars. In contrast to rotation-powered pulsars, the strongly magnetized neutron stars rotate slowly and, therefore, are powered by their magnetic field, see for example \cite{olausen2014mcgill}.  
\section{Magnetized hybrid stars}
In this section we discuss the effects of strong magnetic fields on the global properties of neutron stars, which are subjected to a sharp quark-hadron phase transition in their interior.  Stationary and axi-symmetric stellar models are constructed with the same numerical procedure and mathematical set-up as in \cite{Bocquet:1995je,Chatterjee:2014qsa,Franzon:2015sya}, where  the coupled Einstein-Maxwell equations were solved in a self-consistent way by means of a pseudo-spectral method. 

It was shown in \cite{Chatterjee:2014qsa,Franzon:2015sya} that the leading contribution to the macroscopic properties of strongly magnetized stars, like mass and radius, originates from the pure field contribution to the energy-momentum tensor. In addition, the inclusion of magnetic fields effects in the EoS and the interaction between the magnetic field and matter (the magnetization)  do not affect the stellar structure considerably. In this context, we do not take into consideration the magnetic field effects in our equations of state.

According to \cite{Bocquet:1995je, cardall2001effects}, the magnetic field is generated by the azimuthal component the electromagnetic current 4-vector $j^{u}$:
\begin{equation}
j^{\phi} = \Omega j^{t} + (e+p)f_{0},
\label{current}
\end{equation}
with $j^{t}$ being the time component of the electric current, $\Omega$ the stellar angular velocity, $e$ the energy density and $p$ the isotropic contribution to the pressure. The magnetic stellar models are obtained by assuming a constant current functions $f_{0}$. As shown in \cite{Bocquet:1995je}, other choices for $f_{0}$ different from a constant value are possible, however, they do not alter the conclusions qualitatively. Nevertheless, a more comprehensive study of the field changes and the corresponding variation of current distributions would be very desirable. Such an analysis, however, requires much more insight into the microscopics of the currents in the different hadronic and quark phases and is beyond the scope of this initial discussion of possible observable effects of field decay in highly magnetized stars.
As one can see from Eq.~\ref{current}, for different values of $f_{0}$, the electric current changes and, therefore, the intensity of the magnetic field in the star changes.

\begin{figure}\begin{center}
\includegraphics[width=0.7\textwidth,angle=-90,scale=0.5]{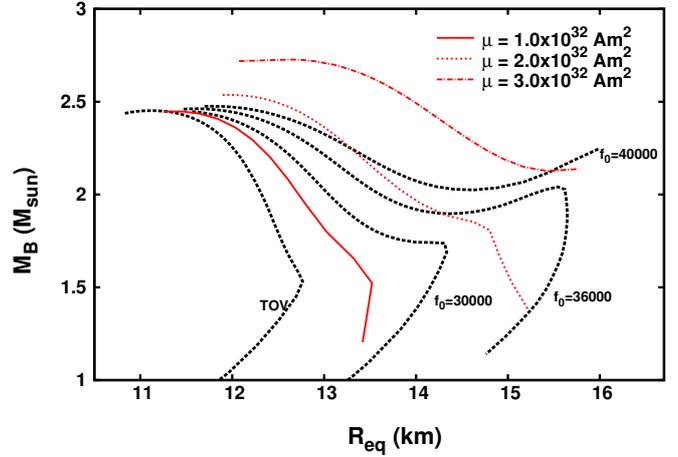}
\caption{\label{mass_radius_bfield} Mass-radius diagram for magnetized models. The calculations were done for different fixed current functions $f_{0}$ and different fixed magnetic dipole moments $\mu$. }
\end{center}
\end {figure}

We show in Fig.~\ref{mass_radius_bfield} the mass-radius diagram for  stars  at different fixed magnetic dipole moments $\mu$ and different current functions $f_{0}$.  
%The magnetic dipole moment $\mu$ is defined as (see Ref.~\cite{Bonazzola:1993zz}):
%\begin{equation}
%\frac{2\mu cos\theta}{r^{3}} = B_{(r)}\mid_{r\rightarrow \infty},
%\label{mm}
%\end{equation}
%which is simply the radial component (the orthonormal one) of the magnetic field of a magnetic dipole seen by an observer at infinity. 
From Fig.~\ref{mass_radius_bfield}, the masses and the radii increase by increasing $\mu$ and $f_{0}$. This is an effect of the Lorentz force which acts outwards and against gravity and, therefore, the stars increase in size and can support more mass.  A star with $M_{B}=2.15\,M_{\odot}$ would be represented by a horizontal line in Fig.~\ref{mass_radius_bfield}, in other words, it corresponds to a set of evolutionary sequences with  smaller magnetic dipole moments. In this case, the magnetic moment loss in  can be related to the change of the magnetic flux  strength and distribution in the star due to ohmic dissipation  \cite{goldreich1992magnetic, heyl1998common}.

The existence of the backbending phenomenon in fast rotating neutron stars are determined by the combination of three quantities: the baryon mass $M_{B}$, the total angular momentum $J$ and the rotational frequency $f$. Minimum values of $M_{B}$ at fixed $f$ with a monotonic behaviour of $M_{B}$ versus $J$ leads to the backbending phenomenon, see e.g. \cite{zdunik2006phase, zdunik2004hyperon}. In parallel to this, we conclude that the backbending in highly magnetized neutron stars depends also on three quantities: the baryon mass $M_{B}$, the magnetic dipole moment $\mu$ and the current function $f_{0}$. In this case, the magnetic dipole loss leads to a quark-hadron phase transition inside the stars, followed by an increase in the  electric current (and, therefore, the magnetic field) through the Eq.~\eqref{current}, which it is related to the change of the type of matter in the star with a different equation of state. In contrast, rotating stars at fixed baryon masses  have their frequency increased by angular moment loss during the backbending epoch.

\begin{figure}\begin{center}
\includegraphics[width=0.7\textwidth,angle=-90,scale=0.5]{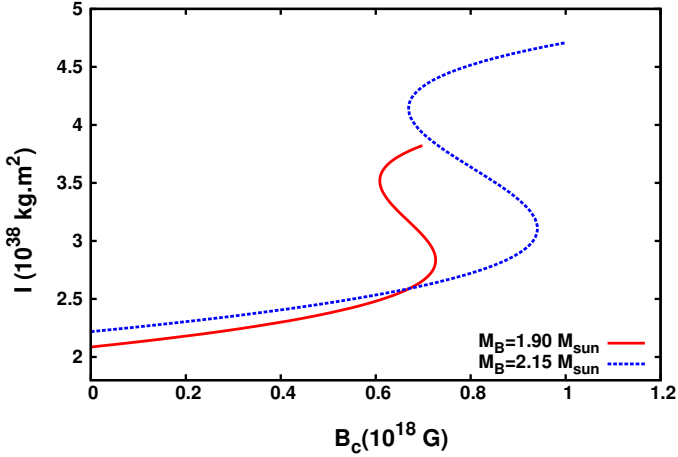}
\caption{\label{iner_bc_maxwell} Relation between the moment of inertia and central magnetic field for magnetized neutron stars with different fixed  baryon masses, 1.90 $\rm{M_{\odot}}$ and 2.15 $\rm{M_{\odot}}$ , respectively.}
\end{center}
\end {figure}

\begin{figure}\begin{center}
\includegraphics[width=0.7\textwidth,angle=-90,scale=0.5]{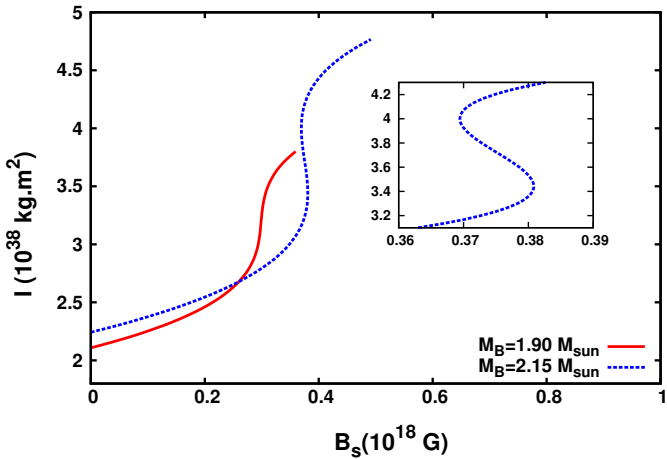}
\caption{\label{iner_bs_maxwell} Same as in Fig.~\ref{iner_bc_maxwell} but as a function of the surface magnetic field. }
\end{center}
\end {figure}

As we already discussed for the case of rotation (but without magnetic field) in Fig.~\ref{iner_omega_maxwell}, the moment of inertia changes
drastically from a constant value in the case of a more realistic treatment. In addition, a slower reduction of the moment of inertia, which is followed by a spin-up of the star, is observed when the equation of state that describes the matter inside these objects includes a strong quark-hadron phase transition. We study the effect of magnetic fields on the moment of inertia $I$ for highly magnetized stars in Fig.~\ref{iner_bc_maxwell} and Fig.~\ref{iner_bs_maxwell}. We present $I$ as a function of the central and surface star magnetic fields, $B_{c}$ and $B_{s}$, respectively. These calculations are done for stars with the same fixed baryon masses as the ones presented in Fig.~\ref{iner_omega_maxwell}.  

From Fig.~\ref{iner_bc_maxwell}, the higher the central magnetic field, the higher the  moment of inertia of the stars. This effect is due to the Lorentz force which allow stars to  support more mass. In addition, the circular equatorial radius of the sequence increases, as can be seen in the mass-radius diagram in Fig.~\ref{mass_radius_bfield} for higher current functions or magnetic dipole moments. According to Fig.~\ref{iner_bc_maxwell}, the maximum central magnetic field reached in stars depends strongly on the stellar mass. For example, a star with $M_{B}=1.90\,\rm{M_{\odot}}$ has a maximum central magnetic field of $\sim$7.0$\times10^{17}$ G, whereas  the star with $M_{B}=2.15\,\rm{M_{\odot}}$ can have a central magnetic field up to 1.0$\times10^{18}$ G. 

If knowing that the magnetic field decays over time, and fixing the baryon mass, each curve depicted in Fig.~\ref{iner_bc_maxwell} represents the time evolution of the stellar  magnetic field and moment of inertia of a different star. In other words, younger stars decrease in size and as the magnetic field decays, the central density increases, see e.g. \cite{Franzon:2015sya}. During this time, these stars might change from a hadronic to a quark phase in the core.  In particular,  the increase of the central magnetic field as shown in Fig.~\ref{iner_bc_maxwell} may represent a signature of a phase transition inside these objects. 
%We call this effect magnetic-up era.  
However, internal magnetic fields cannot be directly constrained by observation. For this reason, we present in Fig.~\ref{iner_bs_maxwell} the same star configurations as shown in Fig.~\ref{iner_bc_maxwell}, but as a function of the (polar) surface magnetic field, which can potentially be observed.    

Looking at the star with fixed baryon mass of 1.90 $\rm{M_{\odot}}$ in Fig.~\ref{iner_bs_maxwell}, one sees that the moment of inertia, which decreases in time, has a slower reduction when the stars passes through the  phase transition.  However, for the star with baryon mass of 2.15 $\rm{M_{\odot}}$, the surface magnetic field in fact increases due to the phase transition. In this case, the moment of inertia as a function of $B_{s}$ exhibits a small ``magnetic-up era", whereas this same effect is much more evident when the central magnetic field $B_{c}$ is considered. 

According to Fig.~\ref{iner_bc_maxwell}, during the ``magnetic-up era", a $M_{B}=2.15\,\rm{M_{\odot}}$ star has increased the value of its central magnetic field by an amount of 0.23$\times 10^{18}$ G, while the surface magnetic field varies from 0.37$\times 10^{18}$ G  to 0.38$\times 10^{18}$ G (see Fig.~\ref{iner_bs_maxwell}). For a star with $M_{B}=1.90\,\rm{M_{\odot}}$, the central magnetic field increases by an amount of 0.1$\times 10^{18}$ G, while the surface magnetic field always decreases. Such effects might be associated with giant flares presented in magnetars \cite{Mallick:2012zq}. In addition, properties of these objects such as neutrino emission and, consequently, the stellar cooling can be strongly affected by this variation in the magnetic field strength. Studies in this line are being performed. 

By increasing the magnetic field strength, the stellar deformation is much more significant. As a result, the shape  of the star will become more elongated and a topological change to a toroidal configuration can take place \cite{cardall2001effects}. However, our current numerical tools do not handle toroidal configuration, setting a limit for the magnetic field strengths that we can obtain within this approach. 

\begin{figure}\begin{center}
% usando estrela no lorene com Mb = 2.15 fixo e corrente j0 = 40000 
\includegraphics[width=1.0\textwidth,angle=-90,scale=0.5]{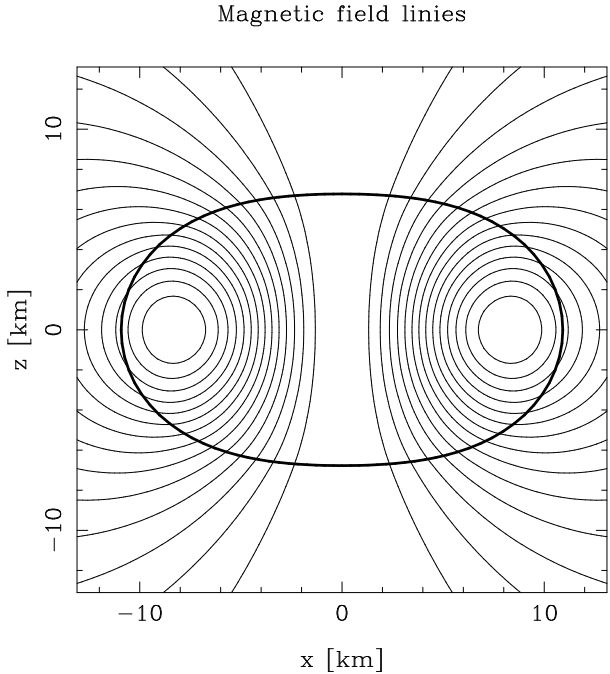}
\caption{\label{deformation}  Magnetic field, i.e $A_{\phi}$, iso-contours measured by the Eulerian observer $\mathcal{O}_{0}$ for a star at fixed baryon mass of 2.15 $\rm{M_{\odot}}$. This star is near the maximum equilibrium configuration achieved by the code with a magnetic dipole moment of $\mu = 2.36\times 10^{32}\rm{\,Am^{2}}$. }
\end{center}
\end {figure}

In order to see the deviation of spherical symmetry due to the anisotropy of the energy-momentum tensor in presence of strong magnetic fields, we show in Fig.~\ref{deformation} a star with $M_{B}=2.15\,\rm{M_{\odot}}$. The corresponding central baryon density is $n=0.45\,\rm{fm^{-3}}$, with a gravitational mass of 1.92 $\rm{M_{\odot}}$. The polar and the central magnetic fields are 4.62$\,\times 10^{17}$ G and 1.03$\,\times 10^{18}$ G, respectively. The ratio between the magnetic pressure and the matter pressure at the center of this star is 0.42. This configuration is quite close to the maximum  magnetic field configuration achieved with the code.      
One sees that the deviation from spherical symmetry is remarkable and it needs to be taken into account while modelling these highly magnetized neutron star. Studies of
relativistic models of magnetized stars with  toroidal magnetic fields
demonstrated also that the deviation from spherical symmetry is important, see \cite{frieben2012equilibrium}. Therefore, a simple Tolman-Oppenheimer-Volkoff general relativist solution \cite{tolman1939static, oppenheimer1939massive} for these stars can not be applied.

\begin{figure}\begin{center}
% usando estrela no lorene com Mb = 2.15 fixo e corrente j0 = 40000 
\includegraphics[width=0.8\textwidth,angle=-90,scale=0.4]{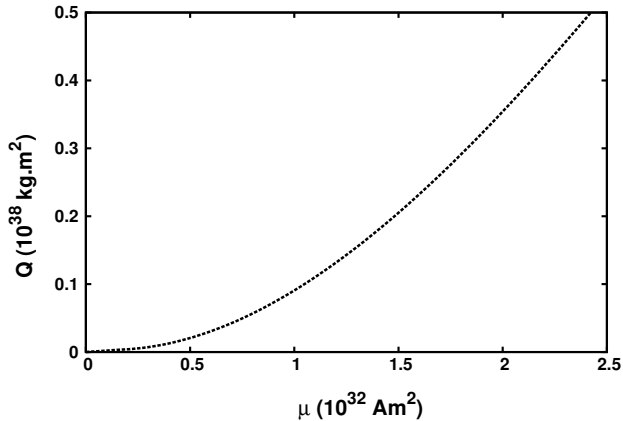}
\caption{\label{quadrupole}  Quadrupole $Q$ as a function of the magnetic dipole moment for a star at fixed baryon mass of 2.15 $\rm{M_{\odot}}$. This same star is depicted in Fig.~\ref{deformation}.}
\end{center}
\end {figure}

The deformation of magnetized neutron stars can also be quantified by their quadrupole moment $Q$ with respect to the rotational axis. In Fig.~\ref{quadrupole}, we show the quadrupole moment Q as a function of the magnetic moment $\mu$ for a star at fixed baryon mass of 2.15 $\rm{M_{\odot}}$. In \cite{bonazzola1996gravitational}, it was found that in a low B-field approximation, $Q$ scales as ${\mu}^2$. From Fig.~\ref{quadrupole}, one sees that indeed the quadrupole moment $Q$ grows parabolically for low polar magnetic fields.
Still according to  \cite{bonazzola1996gravitational},  the gravitational wave (GW) amplitude $h_{0}$ is given by:
\begin{equation}
h_{0}= \frac{6G}{c^{4}}\frac{\Omega^{2}}{D}Q
\label{gw}
\end{equation}
with $G$ being the gravitational constant, $c$ the speed of light,  $D$ the distance of the star  and $\Omega$ the rotational velocity of the star.   

In our models the rotation and magnetic axes are  aligned. In this case, even stars strongly deformed do not emit gravitational radiation. However, an estimate of the strength of gravitational wave (GW) emission can be deduced if we assume that the magnetic axis and the rotation axis are not aligned, as it seems to be the case,  for example, in observable pulsars \cite{bonazzola1996gravitational}.  
This is a good approximation as long as rotational and magnetic field effects do not start to compete in deforming the star. For strong magnetic fields and as long as the rotation frequency is well below the Kepler frequency, this assumption should hold true.

\begin{figure}\begin{center}
% usando estrela no lorene com Mb = 2.15 fixo e corrente j0 = 40000 
\includegraphics[width=0.8\textwidth,angle=-90,scale=0.4]{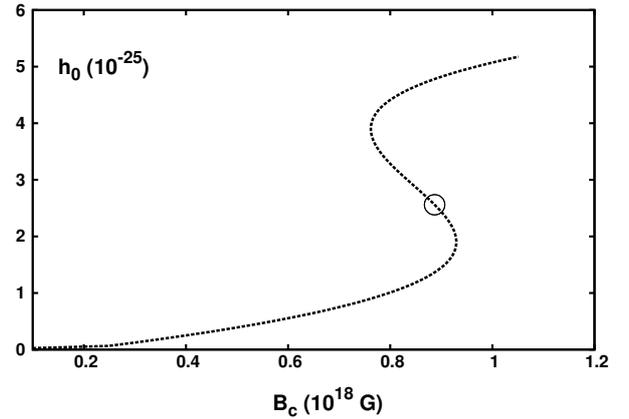}
\caption{\label{h0_bc}  Gravitational wave amplitude as a function of the central magnetic field for the same configurations as in Fig.~\ref{deformation} and Fig.~\ref{quadrupole}} 
\end{center}
\end {figure}

We  use our models to make a crude estimate of the GW strength in highly magnetized neutron stars. Assuming that a star with $M_{B}=2.15\,\rm{M_{\odot}}$ rotates at a  frequency of $f=\Omega/2\pi=$ 1Hz at a distance $D=$10 kpc,  we obtain a gravitational wave amplitude of $h_{0}=$2.56$\times10^{-25}$ (see the black circle in Fig.~\ref{h0_bc}), which could be measured by LIGO and VIRGO interferometric detectors \cite{bonazzola1996gravitational, bonazzola1994astrophysical}.  

%\begin{figure}\begin{center}
% usando estrela no lorene com Mb = 2.15 fixo e corrente j0 = 40000 
%\includegraphics[width=0.8\textwidth,angle=-90,scale=0.4]{h0_quadrupole.eps}
%\input{mr_nocrust_b_magmom}
%\caption{\label{h0}  Gravitational wave amplitude as a function of quadrupole moment $Q$ for the same star as shown in Fig.~\ref{deformation} and Fig.~\ref{quadrupole}. The black circle represents a star with $M_{B}$=2.15 $M_{\odot}$ at a distance of 10 kpc rotating with a frequency of $1$Hz.} 
%\end{center}
%\end {figure}
%In our study, we computed stellar configurations up to this maximum limit,  although the surface magnetic field strengths used in this work lie well above the ones typically observed in magnetars. 

In Fig.~\ref{h0_bc}, we estimate the gravitational wave emission amplitude for different central magnetic fields. As already depicted in the relation between the moment of inertia $I$ and the magnetic field (see Fig.~\ref{iner_bc_maxwell}), we see that the GW amplitude $h_{0}$ can change significantly as the magnetic field decays over time showing a backbending behaviour as the star undergoes a quark-hadron phase transition. According to Fig.~\ref{h0_bc}, the star may have a period of faster reduction in the gravitation waves emission before the magnetic reduces completely. In Fig.~\ref{h0_bc} the black circle represents the same stars as depicted in Fig.~\ref{deformation}. 

\section{conclusions}
In the present work, we studied the effects of quark-hadron phase transition on the structure of neutron stars, considering rotation as well as strong magnetic fields. The stellar models were obtained within a general relativity approach which solves the Einstein-Maxwell equations self-consistently by means of a spectral method.  We employed an equation of state that describes massive hybrid stars composed of nucleons, leptons and quarks $u$, $d$ and $s$. The hadronic phase was described in a relativistic mean field formalism, where many-body forces were taken into consideration. For the quark phase we used the MIT bag model with a vector interaction. As in previous calculations \cite{zdunik2006phase, chubarian1999deconfinement}, we showed that by employing this type of EoS with a sharp phase order transition one obtains the back-bending phenomenon.

The change from a confined to a deconfinement phase at the center of neutron stars leads to a drastic
softening of the equation of state. As a result,  stars at fixed baryon masses decrease their gravitational masses (and the central pressure increases) during the evolutionary sequence. When the softening of the EoS is very pronounced, this leads to a sudden contraction of the neutron star at a critical  angular velocity, which can be observed during the evolution of fast rotating and isolated pulsars through the spin-up era. 
%$This period in the star's life is characterized by the angular moment loss, in other words, the backbending phenomena.
  
For stars at fixed baryon masses of  1.90 and 2.15 $\rm{M_{\odot}}$, we have shown that, as they slow down due to the angular momentum loss through magnetic dipole radiation,  the internal density  increases because of the weakening of the centrifugal force, leading the stars to reach the conditions for a phase transition in their interior. As a result of the phase change, the braking index changes dramatically  around the frequency where the phase transition takes place, showing a divergent point.

We have also studied the effect of a quark-hadron phase transitions in neutron stars with strong magnetic fields.  We concluded that the mass-radius diagram and the shape of the star change significantly with the consistent inclusion of magnetic fields. The mass excess of the star is related to the Lorenz force, which increases with the magnetic field and, therefore,  helps the star to support more mass than in the non-magnetized case. At the same time, the equatorial radii of theses stars increase, becoming much larger than their non-magnetized counterpart.

In parallel to the rotating case, we carried out relativistic calculations of the mass-radius diagram and the moment of inertia for hybrid stars endowed with strong magnetic fields.  We have neglected the effect of the magnetic field in the equation of state and of the magnetization of matter, since it was already shown that these contributions are very small even at large magnetic fields \cite{Franzon:2015sya}. According to our results, in the same way that isolated pulsars have their frequencies increased in presence of a quark-hadron phase transition, the magnetic field can be amplified in highly magnetized hybrid neutron stars.  We performed calculations for stars at two different baryon masses, 1.90 and 2.15 $\rm{M_{\odot}}$, respectively. 

The relation $M_{B}({R_{eq})}$ presents similar behaviour in both rotating (and non-magnetized) and magnetized (and non-rotating) cases. For fast rotating pulsars, it is already known that the backbending phenomenon depends on three quantities: baryon mass $M_{B}$, the angular momentum $J$ and the rotation frequency $f$. For strong magnetized stars, the increasing softening of the EOS leads to inflexions in the diagram $M_{B}({R_{eq})}$ versus $f_{0}$ with a monotonic behaviour at fixed magnetic dipole moments $\mu$. This is a similar condition expected to produce the backbending in isolated and fast rotating pulsars. In this case, the triple values ($M_{B}$, $\mu$, $f_{0}$) corresponds to ($M_{B}$, $J$, $f$).     

The quark-hadron phase transition induces not only structural changes in neutron stars, but can also modify their internal magnetic fields, what can affect, for instance, the thermal evolution processes. Magnetars have been observed to have very strong surface magnetic fields, $B_{s}$.  For this reason, we performed calculations of the moment of inertia as a function of $B_{s}$.  We have shown that the appearance of a ``magnetic-up era"  is strongly related to the stellar mass. For these cases,  we used a purely poloidal magnetic configuration with values for the surface magnetic fields beyond those observed. On the other hand, from virial theorem arguments one can estimate a maximum limit for the magnetic field in the stellar interior and the values considered here agree with realistic situations. 

We have also solved the Einstein-Maxwell equations and
obtained the quadrupole moment $Q$ for slowly rotating magnetized neutron stars. Although these stellar models have rotating and magnetic axes aligned and, therefore, do not emit gravitation waves, we use them to make a crude estimate of the gravitational radiation emitted by a star with $M_{B}=2.15\,\rm{M_{\odot}}$. The value obtained might lead to a detectable signal by VIRGO and LIGO. Note that the gravitational wave amplitude $h_{0}$ can be reduced rapidly in presence of a quark-hadron phase transition (see Fig.\ref{h0_bc}) inside stars with  strong magnetic fields which, again, might be detected by VIRGO and LIGO. 

As shown before in \cite{markey1973adiabatic, tayler1973adiabatic, wright1973pinch, flowers1977evolution},  simple magnetic field configurations composed of purely poloidal or purely toroidal magnetic field configurations are always unstable. In addition,  recent calculations showed that stable equilibrium configurations are possible only with magnetic fields  configurations composed by a poloidal and a toroidal components ~\cite{armaza2015magnetic, braithwaite2004fossil, braithwaite2006stable, akgun2013stability}. 
Finally,  the authors in \cite{goldreich1992magnetic} showed that ohmic decay, ambipolar diffusion, and Hall drift are responsible for the magnetic field decay in isolated neutron stars. In this case, the magnetic field flux might be reduced and change its strength and distribution in the star \cite{goldreich1992magnetic, srinivasan1990novel, heyl1998common}. These points should be addressed in future investigations. 

%In summary, in this study we have modelled highly magnetized hybrid stars with a purely poloidal magnetic field  in a fully general relativity way. A poloidal field geometry is not the most general case, but by using these restrictive assumptions we have seen that the microphysics of a quark-hadron phase transition can potentially affect the macroscopic and observable magnetic fields of such stars.  Additionally, in order to better understand the stars presented here, the thermal evolution of such configurations will be reported in upcoming work. 

 %Our models of rapidly rotating protoneutron stars
%were based on several simplifications, which were neces-
%sary in order to make the problem tractable. In order
%to avoid difficulties and/or ambiguities of the largely un-
%known âreal situationâ, we restricted ourselves to study-
%ing idealized, limiting cases. In many places we introduced
%approximations, which were crucial for making numerical
%calculations feasible.

%In this work, we developed a self-consistent approach to
%determine the structure of neutron stars in strong mag-
%netic fields, relevant for the study of magnetars.
\section*{Acknowledgements}
We gratefully acknowledge the referee  for the valuable comments and suggestions, and Veronica Dexheimer for carefully reading the manuscript. B. Franzon acknowledges support from CNPq/Brazil, DAAD and HGS-HIRe for FAIR. S. Schramm acknowledges support from the HIC for FAIR LOEWE program.  R.O. Gomes would like to thank the fruitful stay at FIAS. This work is partially supported by grant Nr. BEX 14116/13-8 of the PDSE CAPES and Science without Borders programs 
which are an initiative of the Brazilian Government. The authors wish to acknowledge the "NewCompStar" COST Action MP1304.

%%%%%%%%%%%%%%%%%%%%%%%%%%%%%%%%%%%%%%%%%%%%%%%%%%



\bibliographystyle{mnras}
\bibliography{biblio_brak} % if your bibtex file is called example.bib

\begin{thebibliography}{}
\makeatletter
\relax
\def\mn@urlcharsother{\let\do\@makeother \do\$\do\&\do\#\do\^\do\_\do\%\do\~}
\def\mn@doi{\begingroup\mn@urlcharsother \@ifnextchar [ {\mn@doi@}
  {\mn@doi@[]}}
\def\mn@doi@[#1]#2{\def\@tempa{#1}\ifx\@tempa\@empty \href
  {http://dx.doi.org/#2} {doi:#2}\else \href {http://dx.doi.org/#2} {#1}\fi
  \endgroup}
\def\mn@eprint#1#2{\mn@eprint@#1:#2::\@nil}
\def\mn@eprint@arXiv#1{\href {http://arxiv.org/abs/#1} {{\tt arXiv:#1}}}
\def\mn@eprint@dblp#1{\href {http://dblp.uni-trier.de/rec/bibtex/#1.xml}
  {dblp:#1}}
\def\mn@eprint@#1:#2:#3:#4\@nil{\def\@tempa {#1}\def\@tempb {#2}\def\@tempc
  {#3}\ifx \@tempc \@empty \let \@tempc \@tempb \let \@tempb \@tempa \fi \ifx
  \@tempb \@empty \def\@tempb {arXiv}\fi \@ifundefined
  {mn@eprint@\@tempb}{\@tempb:\@tempc}{\expandafter \expandafter \csname
  mn@eprint@\@tempb\endcsname \expandafter{\@tempc}}}

\bibitem[\protect\citeauthoryear{Akg{\"u}n, Reisenegger, Mastrano  \&
  Marchant}{Akg{\"u}n et~al.}{2013}]{akgun2013stability}
Akg{\"u}n T.,  Reisenegger A.,  Mastrano A.,   Marchant P.,  2013, Monthly
  Notices of the Royal Astronomical Society, 433, 2445

\bibitem[\protect\citeauthoryear{Alaverdyan, Alaverdyan  \&
  Chiladze}{Alaverdyan et~al.}{2010}]{Alaverdyan:2010zz}
Alaverdyan A.~G.,  Alaverdyan G.~B.,   Chiladze A.~O.,  2010, \mn@doi [Int. J.
  Mod. Phys.] {10.1142/S0218271810017408}, D19, 1557

\bibitem[\protect\citeauthoryear{Alford, Braby, Paris  \& Reddy}{Alford
  et~al.}{2005}]{Alford:2004pf}
Alford M.,  Braby M.,  Paris M.~W.,   Reddy S.,  2005, \mn@doi [Astrophys. J.]
  {10.1086/430902}, 629, 969

\bibitem[\protect\citeauthoryear{Alvarez-Castillo \& Blaschke}{Alvarez-Castillo
  \& Blaschke}{2015}]{Alvarez-Castillo:2014dva}
Alvarez-Castillo D.~E.,  Blaschke D.,  2015, \mn@doi [Phys. Part. Nucl.]
  {10.1134/S1063779615050032}, 46, 846

\bibitem[\protect\citeauthoryear{Armaza, Reisenegger  \& Valdivia}{Armaza
  et~al.}{2015}]{armaza2015magnetic}
Armaza C.,  Reisenegger A.,   Valdivia J.~A.,  2015, The Astrophysical Journal,
  802, 121

\bibitem[\protect\citeauthoryear{Ayvazyan, Colucci, Rischke  \&
  Sedrakian}{Ayvazyan et~al.}{2013}]{Ayvazyan:2013cva}
Ayvazyan N.~S.,  Colucci G.,  Rischke D.~H.,   Sedrakian A.,  2013, \mn@doi
  [Astron. Astrophys.] {10.1051/0004-6361/201322484}, 559, A118

\bibitem[\protect\citeauthoryear{Bhattacharyya, Mishustin  \&
  Greiner}{Bhattacharyya et~al.}{2010}]{Bhattacharyya:2009fg}
Bhattacharyya A.,  Mishustin I.~N.,   Greiner W.,  2010, \mn@doi [J. Phys.]
  {10.1088/0954-3899/37/2/025201}, G37, 025201

\bibitem[\protect\citeauthoryear{Blandford \& Romani}{Blandford \&
  Romani}{1988}]{blandford1988interpretation}
Blandford R.~D.,  Romani R.~W.,  1988, Monthly Notices of the Royal
  Astronomical Society, 234, 57P

\bibitem[\protect\citeauthoryear{Bocquet, Bonazzola, Gourgoulhon  \&
  Novak}{Bocquet et~al.}{1995}]{Bocquet:1995je}
Bocquet M.,  Bonazzola S.,  Gourgoulhon E.,   Novak J.,  1995, Astron.
  Astrophys., 301, 757

\bibitem[\protect\citeauthoryear{Bombaci, Lugones  \& Vidana}{Bombaci
  et~al.}{2007}]{Bombaci:2006cs}
Bombaci I.,  Lugones G.,   Vidana I.,  2007, \mn@doi [Astron. Astrophys.]
  {10.1051/0004-6361:20065259}, 462, 1017

\bibitem[\protect\citeauthoryear{Bombaci, Logoteta, Panda, Providencia  \&
  Vidana}{Bombaci et~al.}{2009}]{Bombaci:2009jt}
Bombaci I.,  Logoteta D.,  Panda P.~K.,  Providencia C.,   Vidana I.,  2009,
  \mn@doi [Phys. Lett.] {10.1016/j.physletb.2009.09.039}, B680, 448

\bibitem[\protect\citeauthoryear{Bonazzola \& Gourgoulhon}{Bonazzola \&
  Gourgoulhon}{1996}]{bonazzola1996gravitational}
Bonazzola S.,  Gourgoulhon E.,  1996, Astron. Astrophys., 312, 675

\bibitem[\protect\citeauthoryear{Bonazzola \& Marck}{Bonazzola \&
  Marck}{1994}]{bonazzola1994astrophysical}
Bonazzola S.,  Marck J.,  1994, Annual Review of Nuclear and Particle Science,
  44, 655

\bibitem[\protect\citeauthoryear{Bonazzola, Gourgoulhon, Salgado  \&
  Marck}{Bonazzola et~al.}{1993}]{Bonazzola:1993zz}
Bonazzola S.,  Gourgoulhon E.,  Salgado M.,   Marck J.,  1993, Astronomy and
  Astrophysics, 278, 421

\bibitem[\protect\citeauthoryear{Braithwaite \& Nordlund}{Braithwaite \&
  Nordlund}{2006}]{braithwaite2006stable}
Braithwaite J.,  Nordlund {\AA}.,  2006, Astronomy \& Astrophysics, 450, 1077

\bibitem[\protect\citeauthoryear{Braithwaite \& Spruit}{Braithwaite \&
  Spruit}{2004}]{braithwaite2004fossil}
Braithwaite J.,  Spruit H.~C.,  2004, Nature, 431, 819

\bibitem[\protect\citeauthoryear{Brillante \& Mishustin}{Brillante \&
  Mishustin}{2014}]{Brillante:2014lwa}
Brillante A.,  Mishustin I.~N.,  2014, \mn@doi [Europhys. Lett.]
  {10.1209/0295-5075/105/39001}, 105, 39001

\bibitem[\protect\citeauthoryear{Cardall, Prakash  \& Lattimer}{Cardall
  et~al.}{2001}]{cardall2001effects}
Cardall C.~Y.,  Prakash M.,   Lattimer J.~M.,  2001, The Astrophysical Journal,
  554, 322

\bibitem[\protect\citeauthoryear{Chatterjee, Elghozi, Novak  \&
  Oertel}{Chatterjee et~al.}{2015}]{Chatterjee:2014qsa}
Chatterjee D.,  Elghozi T.,  Novak J.,   Oertel M.,  2015, \mn@doi [Mon. Not.
  Roy. Astron. Soc.] {10.1093/mnras/stu2706}, 447, 3785

\bibitem[\protect\citeauthoryear{Cheng, Yuan  \& Zhang}{Cheng
  et~al.}{2002}]{cheng2002phase}
Cheng K.,  Yuan Y.,   Zhang J.,  2002, The Astrophysical Journal, 564, 909

\bibitem[\protect\citeauthoryear{Chubarian, Grigorian, Poghosyan  \&
  Blaschke}{Chubarian et~al.}{2000}]{chubarian1999deconfinement}
Chubarian E.,  Grigorian H.,  Poghosyan G.~S.,   Blaschke D.,  2000, Astron.
  Astrophys., 357, 968

\bibitem[\protect\citeauthoryear{Contrera, Spinella, Orsaria  \&
  Weber}{Contrera et~al.}{2014}]{contrera2014hadron}
Contrera G.~A.,  Spinella W.,  Orsaria M.,   Weber F.,  2014, arXiv preprint
  arXiv:1403.7415

\bibitem[\protect\citeauthoryear{Cook, Shapiro  \& Teukolsky}{Cook
  et~al.}{1992}]{cook1992spin}
Cook G.~B.,  Shapiro S.~L.,   Teukolsky S.~A.,  1992, The Astrophysical
  Journal, 398, 203

\bibitem[\protect\citeauthoryear{Denke \& Pinto}{Denke \&
  Pinto}{2013}]{denke2013influence}
Denke R.~Z.,  Pinto M.~B.,  2013, Physical Review D, 88, 056008

\bibitem[\protect\citeauthoryear{Dexheimer, Schramm  \& Stone}{Dexheimer
  et~al.}{2012}]{Dexheimer:2012mm}
Dexheimer V.,  Schramm S.,   Stone J.,  2012, PoS, NICXII, 101

\bibitem[\protect\citeauthoryear{Dexheimer, Negreiros  \& Schramm}{Dexheimer
  et~al.}{2015}]{Dexheimer:2014pea}
Dexheimer V.,  Negreiros R.,   Schramm S.,  2015, \mn@doi [Phys. Rev.]
  {10.1103/PhysRevC.91.055808}, C91, 055808

\bibitem[\protect\citeauthoryear{Duncan \& Thompson}{Duncan \&
  Thompson}{1992}]{Duncan:1992hi}
Duncan R.~C.,  Thompson C.,  1992, \mn@doi [Astrophys. J.] {10.1086/186413},
  392, L9

\bibitem[\protect\citeauthoryear{Farhi \& Jaffe}{Farhi \&
  Jaffe}{1984}]{Farhi:1984qu}
Farhi E.,  Jaffe R.~L.,  1984, \mn@doi [Phys. Rev.] {10.1103/PhysRevD.30.2379},
  D30, 2379

\bibitem[\protect\citeauthoryear{Ferrari \& Ruffini}{Ferrari \&
  Ruffini}{1969}]{ferrari1969theoretical}
Ferrari A.,  Ruffini R.,  1969, The Astrophysical Journal, 158, L71

\bibitem[\protect\citeauthoryear{Ferrer, de La~Incera, Keith, Portillo  \&
  Springsteen}{Ferrer et~al.}{2010}]{ferrer2010equation}
Ferrer E.~J.,  de La~Incera V.,  Keith J.~P.,  Portillo I.,   Springsteen
  P.~L.,  2010, Physical Review C, 82, 065802

\bibitem[\protect\citeauthoryear{Flowers \& Ruderman}{Flowers \&
  Ruderman}{1977}]{flowers1977evolution}
Flowers E.,  Ruderman M.~A.,  1977, The Astrophysical Journal, 215, 302

\bibitem[\protect\citeauthoryear{Fraga, Pisarski  \& Schaffner-Bielich}{Fraga
  et~al.}{2001}]{Fraga:2001id}
Fraga E.~S.,  Pisarski R.~D.,   Schaffner-Bielich J.,  2001, \mn@doi [Phys.
  Rev.] {10.1103/PhysRevD.63.121702}, D63, 121702

\bibitem[\protect\citeauthoryear{Fraga, Kurkela  \& Vuorinen}{Fraga
  et~al.}{2014}]{Fraga:2013qra}
Fraga E.~S.,  Kurkela A.,   Vuorinen A.,  2014, \mn@doi [Astrophys. J.]
  {10.1088/2041-8205/781/2/L25}, 781, L25

\bibitem[\protect\citeauthoryear{Franzon \& Schramm}{Franzon \&
  Schramm}{2015}]{Franzon:2015gda}
Franzon B.,  Schramm S.,  2015, \mn@doi [Phys. Rev.]
  {10.1103/PhysRevD.92.083006}, D92, 083006

\bibitem[\protect\citeauthoryear{Franzon, Dexheimer  \& Schramm}{Franzon
  et~al.}{2015}]{Franzon:2015sya}
Franzon B.,  Dexheimer V.,   Schramm S.,  2015, \mn@doi [Mon. Not. Roy. Astron.
  Soc.] {10.1093/mnras/stv2606}, 456, 2937

\bibitem[\protect\citeauthoryear{Frieben \& Rezzolla}{Frieben \&
  Rezzolla}{2012}]{frieben2012equilibrium}
Frieben J.,  Rezzolla L.,  2012, Monthly Notices of the Royal Astronomical
  Society, 427, 3406

\bibitem[\protect\citeauthoryear{Friedman \& Ipser}{Friedman \&
  Ipser}{1992}]{friedman1992rapidly}
Friedman J.~L.,  Ipser J.~R.,  1992, Philosophical Transactions of the Royal
  Society of London A: Mathematical, Physical and Engineering Sciences, 340,
  391

\bibitem[\protect\citeauthoryear{Fushiki, Gudmundsson  \& Pethick}{Fushiki
  et~al.}{1989}]{fushiki1989surface}
Fushiki I.,  Gudmundsson E.,   Pethick C.,  1989, The Astrophysical Journal,
  342, 958

\bibitem[\protect\citeauthoryear{Gao, Li, Wang, Yuan, Peng  \& Du}{Gao
  et~al.}{2015}]{gao2015constraining}
Gao Z.,  Li X.,  Wang N.,  Yuan J.,  Peng Q.,   Du Y.,  2015, arXiv preprint
  arXiv:1505.07013

\bibitem[\protect\citeauthoryear{Garcia \& Pinto}{Garcia \&
  Pinto}{2013}]{Garcia:2013eaa}
Garcia A.~F.,  Pinto M.~B.,  2013, \mn@doi [Phys. Rev.]
  {10.1103/PhysRevC.88.025207}, C88, 025207

\bibitem[\protect\citeauthoryear{Glendenning}{Glendenning}{2012}]{glendenning2012compact}
Glendenning N.~K.,  2012, Compact stars: Nuclear physics, particle physics and
  general relativity.
Springer Science \& Business Media

\bibitem[\protect\citeauthoryear{Glendenning, Pei  \& Weber}{Glendenning
  et~al.}{1997}]{glendenning1997signal}
Glendenning N.~K.,  Pei S.,   Weber F.,  1997, Physical Review Letters, 79,
  1603

\bibitem[\protect\citeauthoryear{Gold}{Gold}{1975}]{gold1975rotating}
Gold T.,  1975, Neutron stars, black holes, and binary X-ray sources, 48, 354

\bibitem[\protect\citeauthoryear{Goldreich \& Reisenegger}{Goldreich \&
  Reisenegger}{1992}]{goldreich1992magnetic}
Goldreich P.,  Reisenegger A.,  1992, The Astrophysical Journal, 395, 250

\bibitem[\protect\citeauthoryear{Gomes, Dexheimer, Schramm  \&
  Vasconcellos}{Gomes et~al.}{2015}]{Gomes:2014aka}
Gomes R.~O.,  Dexheimer V.,  Schramm S.,   Vasconcellos C. A.~Z.,  2015,
  \mn@doi [Astrophys. J.] {10.1088/0004-637X/808/1/8}, 808, 8

\bibitem[\protect\citeauthoryear{Gourgoulhon}{Gourgoulhon}{2012}]{gourgoulhon20123+}
Gourgoulhon E.,  2012, 3+ 1 formalism in general relativity: bases of numerical
  relativity.
 Vol. 846, Springer Science \& Business Media

\bibitem[\protect\citeauthoryear{Haensel, Zdunik, Bejger  \& Lattimer}{Haensel
  et~al.}{2009}]{haensel2009keplerian}
Haensel P.,  Zdunik J.,  Bejger M.,   Lattimer J.,  2009, Astronomy \&
  Astrophysics, 502, 605

\bibitem[\protect\citeauthoryear{Hamil, Stone, Urbanec  \& Urbancová}{Hamil
  et~al.}{2015}]{Hamil:2015hqa}
Hamil O.,  Stone J.,  Urbanec M.,   Urbancová G.,  2015, \mn@doi [Phys. Rev.]
  {10.1103/PhysRevD.91.063007}, D91, 063007

\bibitem[\protect\citeauthoryear{Heiselberg \& Hjorth-Jensen}{Heiselberg \&
  Hjorth-Jensen}{1998}]{heiselberg1998phase}
Heiselberg H.,  Hjorth-Jensen M.,  1998, Physical Review letters, 80, 5485

\bibitem[\protect\citeauthoryear{Hempel, Pagliara  \& Schaffner-Bielich}{Hempel
  et~al.}{2009}]{Hempel:2009vp}
Hempel M.,  Pagliara G.,   Schaffner-Bielich J.,  2009, \mn@doi [Phys. Rev.]
  {10.1103/PhysRevD.80.125014}, D80, 125014

\bibitem[\protect\citeauthoryear{Heyl \& Kulkarni}{Heyl \&
  Kulkarni}{1998}]{heyl1998common}
Heyl J.~S.,  Kulkarni S.,  1998, The Astrophysical Journal Letters, 506, L61

\bibitem[\protect\citeauthoryear{Kl{\"a}hn \& Fischer}{Kl{\"a}hn \&
  Fischer}{2015}]{klahn2015vector}
Kl{\"a}hn T.,  Fischer T.,  2015, The Astrophysical Journal, 810, 134

\bibitem[\protect\citeauthoryear{Komatsu, Eriguchi  \& Hachisu}{Komatsu
  et~al.}{1989}]{komatsu1989rapidly}
Komatsu H.,  Eriguchi Y.,   Hachisu I.,  1989, Monthly Notices of the Royal
  Astronomical Society, 237, 355

\bibitem[\protect\citeauthoryear{Lai \& Shapiro}{Lai \&
  Shapiro}{1991}]{lai1991cold}
Lai D.,  Shapiro S.~L.,  1991, The Astrophysical Journal, 383, 745

\bibitem[\protect\citeauthoryear{Lenzi \& Lugones}{Lenzi \&
  Lugones}{2012}]{Lenzi:2012xz}
Lenzi C.~H.,  Lugones G.,  2012, \mn@doi [Astrophys. J.]
  {10.1088/0004-637X/759/1/57}, 759, 57

\bibitem[\protect\citeauthoryear{Lugones, Grunfeld  \& Al~Ajmi}{Lugones
  et~al.}{2013}]{Lugones:2013ema}
Lugones G.,  Grunfeld A.~G.,   Al~Ajmi M.,  2013, \mn@doi [Phys. Rev.]
  {10.1103/PhysRevC.88.045803}, C88, 045803

\bibitem[\protect\citeauthoryear{Mallick \& Sahu}{Mallick \&
  Sahu}{2014}]{Mallick:2012zq}
Mallick R.,  Sahu P.~K.,  2014, \mn@doi [Nucl. Phys.]
  {10.1016/j.nuclphysa.2013.11.009}, A921, 96

\bibitem[\protect\citeauthoryear{Manchester, Durdin  \& Newton}{Manchester
  et~al.}{1985}]{manchester1985second}
Manchester R.,  Durdin J.,   Newton L.,  1985

\bibitem[\protect\citeauthoryear{Markey \& Tayler}{Markey \&
  Tayler}{1973}]{markey1973adiabatic}
Markey P.,  Tayler R.,  1973, Monthly Notices of the Royal Astronomical
  Society, 163, 77

\bibitem[\protect\citeauthoryear{Melatos}{Melatos}{1999}]{melatos1999bumpy}
Melatos A.,  1999, The Astrophysical Journal Letters, 519, L77

\bibitem[\protect\citeauthoryear{Menezes, Pinto, Castro, Costa  \&
  Provid{\^e}ncia}{Menezes et~al.}{2014}]{menezes2014repulsive}
Menezes D.~P.,  Pinto M.~B.,  Castro L.~B.,  Costa P.,   Provid{\^e}ncia C.,
  2014, Physical Review C, 89, 055207

\bibitem[\protect\citeauthoryear{Olausen \& Kaspi}{Olausen \&
  Kaspi}{2014}]{olausen2014mcgill}
Olausen S.,  Kaspi V.,  2014, The Astrophysical Journal Supplement Series, 212,
  6

\bibitem[\protect\citeauthoryear{Oppenheimer \& Volkoff}{Oppenheimer \&
  Volkoff}{1939}]{oppenheimer1939massive}
Oppenheimer J.~R.,  Volkoff G.~M.,  1939, Physical Review, 55, 374

\bibitem[\protect\citeauthoryear{Ostriker \& Gunn}{Ostriker \&
  Gunn}{1969}]{ostriker1969nature}
Ostriker J.,  Gunn J.,  1969, The Astrophysical Journal, 157, 1395

\bibitem[\protect\citeauthoryear{Pacini}{Pacini}{1967}]{pacini1967energy}
Pacini F.,  1967, Nature, 216, 567

\bibitem[\protect\citeauthoryear{Pacini}{Pacini}{1968}]{pacini1968rotating}
Pacini F.,  1968, Nature, 219, 145

\bibitem[\protect\citeauthoryear{Paczynski}{Paczynski}{1992}]{paczynski1992gb}
Paczynski B.,  1992, Acta Astronomica, 42, 145

\bibitem[\protect\citeauthoryear{Ranea-Sandoval, Han, Orsaria, Contrera, Weber
  \& Alford}{Ranea-Sandoval et~al.}{2015}]{ranea2015constant}
Ranea-Sandoval I.~F.,  Han S.,  Orsaria M.~G.,  Contrera G.~A.,  Weber F.,
  Alford M.~G.,  2015, arXiv preprint arXiv:1512.09183

\bibitem[\protect\citeauthoryear{Restrepo, Macias, Pinto  \& Ferrari}{Restrepo
  et~al.}{2015}]{Restrepo:2014fna}
Restrepo T.~E.,  Macias J.~C.,  Pinto M.~B.,   Ferrari G.~N.,  2015, \mn@doi
  [Phys. Rev.] {10.1103/PhysRevD.91.065017}, D91, 065017

\bibitem[\protect\citeauthoryear{Salgado, Bonazzola, Gourgoulhon  \&
  Haensel}{Salgado et~al.}{1994}]{salgado1994high}
Salgado M.,  Bonazzola S.,  Gourgoulhon E.,   Haensel P.,  1994, Astronomy and
  Astrophysics, 291, 155

\bibitem[\protect\citeauthoryear{Shao, Colonna, Di~Toro, Liu  \& Matera}{Shao
  et~al.}{2012}]{Shao:2012tu}
Shao G.~Y.,  Colonna M.,  Di~Toro M.,  Liu B.,   Matera F.,  2012, \mn@doi
  [Phys. Rev.] {10.1103/PhysRevD.85.114017}, D85, 114017

\bibitem[\protect\citeauthoryear{Shao, Colonna, Di~Toro, Liu  \& Liu}{Shao
  et~al.}{2013}]{Shao:2013toa}
Shao G.~Y.,  Colonna M.,  Di~Toro M.,  Liu Y.~X.,   Liu B.,  2013, \mn@doi
  [Phys. Rev.] {10.1103/PhysRevD.87.096012}, D87, 096012

\bibitem[\protect\citeauthoryear{Shapiro \& Teukolsky}{Shapiro \&
  Teukolsky}{2008}]{shapiro2008black}
Shapiro S.~L.,  Teukolsky S.~A.,  2008, Black holes, white dwarfs and neutron
  stars: the physics of compact objects

\bibitem[\protect\citeauthoryear{Spyrou \& Stergioulas}{Spyrou \&
  Stergioulas}{2002}]{spyrou2002spin}
Spyrou N.,  Stergioulas N.,  2002, Astronomy \& Astrophysics, 395, 151

\bibitem[\protect\citeauthoryear{Srinivasan, Bhattacharya, Muslimov  \&
  Tsygan}{Srinivasan et~al.}{1990}]{srinivasan1990novel}
Srinivasan G.,  Bhattacharya D.,  Muslimov A.,   Tsygan A.,  1990, Current
  Science, 59, 31

\bibitem[\protect\citeauthoryear{Stergioulas \& Friedman}{Stergioulas \&
  Friedman}{1994}]{stergioulas1994comparing}
Stergioulas N.,  Friedman J.~L.,  1994, arXiv preprint astro-ph/9411032

\bibitem[\protect\citeauthoryear{Tayler}{Tayler}{1973}]{tayler1973adiabatic}
Tayler R.,  1973, Monthly Notices of the Royal Astronomical Society, 161, 365

\bibitem[\protect\citeauthoryear{Thompson \& Duncan}{Thompson \&
  Duncan}{1993}]{Thompson:1993hn}
Thompson C.,  Duncan R.~C.,  1993, \mn@doi [Astrophys. J.] {10.1086/172580},
  408, 194

\bibitem[\protect\citeauthoryear{Thompson \& Duncan}{Thompson \&
  Duncan}{1996}]{Thompson:1996pe}
Thompson C.,  Duncan R.~C.,  1996, \mn@doi [Astrophys. J.] {10.1086/178147},
  473, 322

\bibitem[\protect\citeauthoryear{Tolman}{Tolman}{1939}]{tolman1939static}
Tolman R.~C.,  1939, Physical Review, 55, 364

\bibitem[\protect\citeauthoryear{Weber, Glendenning  \& Pei}{Weber
  et~al.}{1997}]{weber1997signal}
Weber F.,  Glendenning N.~K.,   Pei S.,  1997, arXiv preprint astro-ph/9705202

\bibitem[\protect\citeauthoryear{Weissenborn, Sagert, Pagliara, Hempel  \&
  Schaffner-Bielich}{Weissenborn et~al.}{2011}]{Weissenborn:2011qu}
Weissenborn S.,  Sagert I.,  Pagliara G.,  Hempel M.,   Schaffner-Bielich J.,
  2011, \mn@doi [Astrophys. J.] {10.1088/2041-8205/740/1/L14}, 740, L14

\bibitem[\protect\citeauthoryear{Wright}{Wright}{1973}]{wright1973pinch}
Wright G.,  1973, Monthly Notices of the Royal Astronomical Society, 162, 339

\bibitem[\protect\citeauthoryear{Yasutake \& Kashiwa}{Yasutake \&
  Kashiwa}{2009}]{Yasutake:2009kj}
Yasutake N.,  Kashiwa K.,  2009, \mn@doi [Phys. Rev.]
  {10.1103/PhysRevD.79.043012}, D79, 043012

\bibitem[\protect\citeauthoryear{Yasutake, Burgio  \& Schulze}{Yasutake
  et~al.}{2011}]{Yasutake:2010eq}
Yasutake N.,  Burgio G.~F.,   Schulze H.~J.,  2011, \mn@doi [Phys. Atom. Nucl.]
  {10.1134/S1063778811100073}, 74, 1502

\bibitem[\protect\citeauthoryear{Zdunik, Haensel, Gourgoulhon  \&
  Bejger}{Zdunik et~al.}{2004}]{zdunik2004hyperon}
Zdunik J.~L.,  Haensel P.,  Gourgoulhon E.,   Bejger M.,  2004, Astronomy \&
  Astrophysics, 416, 1013

\bibitem[\protect\citeauthoryear{Zdunik, Bejger, Haensel  \&
  Gourgoulhon}{Zdunik et~al.}{2006}]{zdunik2006phase}
Zdunik J.,  Bejger M.,  Haensel P.,   Gourgoulhon E.,  2006, Astronomy \&
  Astrophysics, 450, 747

\bibitem[\protect\citeauthoryear{Zdunik, Bejger, Haensel  \&
  Gourgoulhon}{Zdunik et~al.}{2008}]{zdunik2008strong}
Zdunik J.,  Bejger M.,  Haensel P.,   Gourgoulhon E.,  2008, Astronomy \&
  Astrophysics, 479, 515

\bibitem[\protect\citeauthoryear{de Carvalho, Negreiros, Orsaria, Contrera,
  Weber  \& Spinella}{de~Carvalho et~al.}{2015}]{deCarvalho:2015lpa}
de Carvalho S.~M.,  Negreiros R.,  Orsaria M.,  Contrera G.~A.,  Weber F.,
  Spinella W.,  2015, \mn@doi [Phys. Rev.] {10.1103/PhysRevC.92.035810}, C92,
  035810

\makeatother
\end{thebibliography}

% Don't change these lines
\bsp	% typesetting comment
\label{lastpage}
\end{document}